\documentclass[aps,prl,reprint,showpacs,amsmath,amssymb,superscriptaddress,twocolumn]{revtex4-1}
\usepackage[normalem]{ulem}
\usepackage{dcolumn}
\usepackage{bm}
\usepackage{amssymb}
\usepackage{color}
\usepackage{subfigure}
\usepackage[latin1]{inputenc}
\usepackage{graphicx}
\usepackage{latexsym}
\usepackage{amsfonts}
\usepackage{amsmath}
\usepackage{textcomp}

\newcommand{\be}{\begin{equation}}
\newcommand{\ee}{\end{equation}}
\newcommand{\bea}{\begin{eqnarray}}
\newcommand{\eea}{\end{eqnarray}}
\newcommand{\bse}{\begin{subequations}}
\newcommand{\ese}{\end{subequations}}

\newcommand{\comment}[1]{}

\begin{document}

%Multiple backflows from surface waves over dense polymer brushes/ 
\title{Surface wave excitations and backflow effect over dense polymer brushes}

\author{Sofia Biagi}
\email{sofia.biagi@ujf-grenoble.fr}
\affiliation{Universit\'e Grenoble Alpes/CNRS UMR 5588, LIPhy, 38041 Grenoble, France}
\affiliation{Dipartimento di Fisica, Sapienza-Universit\'a di Roma, Piazzale A. Moro 5, 00185 Roma, Italy}
\author{Lorenzo Rovigatti}
%\email{lorenzo.rovigatti@uniroma1.it}
\affiliation{Dipartimento di Fisica, Sapienza-Universit\'a di Roma, Piazzale A. Moro 5, 00185 Roma, Italy}
\affiliation{Faculty of Physics, University of Vienna, Boltzmanngasse 5, A-1090 Vienna, Austria}
\author{Francesco Sciortino}
%\email{francesco.sciortino@uniroma1.it}
\affiliation{Dipartimento di Fisica, Sapienza-Universit\'a di Roma, Piazzale A. Moro 5, 00185 Roma, Italy}
\affiliation{Istituto Sistemi Complessi (ISC), Via dei Taurini 19, 00185 Roma, Italy}
\author{Chaouqi Misbah}
%\email{chaouqi.misbah@ujf-grenoble.fr}
\affiliation{Universit\'e Grenoble Alpes/CNRS UMR 5588, LIPhy, 38041 Grenoble, France}

\date{\today}

\begin{abstract}
Polymer brushes are increasingly
used to tailor surface physicochemistry for various applications such as wetting, adhesion
of biological objects, implantable devices, etc.
We perform Dissipative Particle Dynamics simulations to study the behavior of dense polymer brushes under flow in a slit-pore channel. 
We discover that the system displays flow inversion at the brush interface for {\it several} disconnected ranges of the imposed flow. We associate such phenomenon to collective polymer dynamics: 
a wave propagating on the brush surface. The relation between    the wavelength,  the amplitude and the propagation speed of the flow-generated wave 
is consistent with the solution of the Stokes equations 
when an imposed traveling wave is assumed as boundary condition 
(the famous Taylor's swimmer). 
%We  quantify the inversion interpreting the traveling wave as boundary condition for the Stokes equation and the backflow as a propulsion that cannot be expressed by the fixed brush and is rather transmitted to the fluid./...the inversion within a paradigm of self-propelling microswimmers.
%of active pusher microswimmer.
\end{abstract}

\maketitle

Polymer brushes are passive media
whose great variety allows for a rich range of applications. Brushes with different mechanical properties 
can be created by grafting 
simple polymers,  block copolymers or polymer stars
to a solid substrates or to an interface between two liquids. %(as vesicles)  
They are exploited for colloid stabilization \cite{pincus,milchev,coll_stabil_pH}, as lubricant layers \cite{polyelectrolyte_brush,klein_1994,klein_synovial}, as adhesion regulators \cite{adhesion_2013,adhesion_2011} and for
biomedical technological applications. A holdover interest in these systems  
is indeed motivated by the discovery that 
the inner
surface of various mammalian organs is decorated by densily grafted macromolecules. For example, 
the lumen of blood vessels is protected by a hundred nanometers thick polymer brush mainly made of glucose and its compounds. Such an heterogeneous network is called 
\textquotedblleft glycocalyx\textquotedblright ~\cite{glycocalyx_bio,Reitsma_Slaaf_Vink_2007}.  
Research on the glycocalyx dynamics is central for a complete understanding of the blood circulatory system and, hopefully, to shed light on the initial stages of diseases like thrombosis and atherosclerosis ~\cite{glyco_vascular_disease}.  Understanding
glycocalyx is also relevant to  the polymeric coatings of lungs, small intestine and uterus~\cite{glyco_lung,glyco_uterus,Rubinstein_mucus}.

Most theoretical studies have attempted to model  polymer brushes as porous media
described by Brinkman-type equations ~\cite{secomb_2001} or as elastic media composed of very rigid fibers ~\cite{Weinbaum}. 
%However, recent experimental measurements on artificial glycocalyx under flow have brought to the scientific community attention peculiar phenomena still waiting for an exhaustive explanation ~\cite{L&L}.
%Here, we aim to contribute to the field with an analysis of a polymer brush whose features approach the glycocalyx system and for which the set up conditions recall the microcirculation frame.
In this Letter,  we report an analysis of a polymer brush whose features approach the glycocalyx system and for which the set up conditions recall the microcirculation frame~\cite{L&L}.
We focus on dense polymer brushes under flow, in which the brush is modeled as a
collection of individual polymers. We implement a Dissipative Particle Dynamics (DPD) code~\cite{Hoogerbrugge_Koelman,bridging_the_gap,Espanol_Warren,stripes} with explicit solvent and numerically analyse the dynamics of a linear flexible homo-disperse polymer brush subdued to a simple liquid parabolic flow in a slit-pore geometry.   
The coarse-grained DPD procedure applies to both  solvent molecules and polymer monomers, 
offering the possibility to (i) reproduce  hydrodynamic interactions while retaining a detailed view of the brush dynamics on the scale of the coarse-grained monomers; (ii)  access both  the polymer dynamics, influenced by the imposed flow, and  the flow field, perturbed by the presence and motion of the brush.

Recent  studies of polymer brushes under flow~\cite{Leonforte,Muller_Pastorino,Deng}
 have highlighted an unexpected behavior in the velocity 
profile in the vicinity of the brush surface. These studies have reported that the velocity field reverses 
on increasing the flow field and have tentatively associated such result to the peculiar dynamics of the
single polymer undergoing a cyclic motion of stretching, elongation and recoiling.  Stimulated by
these studies we have undertaken  a numerical investigation exploring a very large range of imposed flow,
aiming at quantifying the conditions under which the physical properties of the brush couple with the
hydrodynamic properties of the solvent to produce flow inversion.  As presented below,
we discover (i) that flow inversion appears in distinct regions of imposed flow values;
(ii) that every time flow inversion is observed a surface wave appears, stressing that such backflow is
strongly associated to a collective (as opposed to single) polymer dynamics; (iii) that the 
wave properties  are consistent with  predictions derived by Taylor in his seminal study on pusher
microswimmers~\cite{Taylor}.  Thus, our work presents a new interpretation for the flow inversion phenomenon and  provide a novel connection between two separated fields: polymer brushes under flow and microswimmers.

\begin{figure}[ht]
\centering%
\includegraphics[width=8cm]{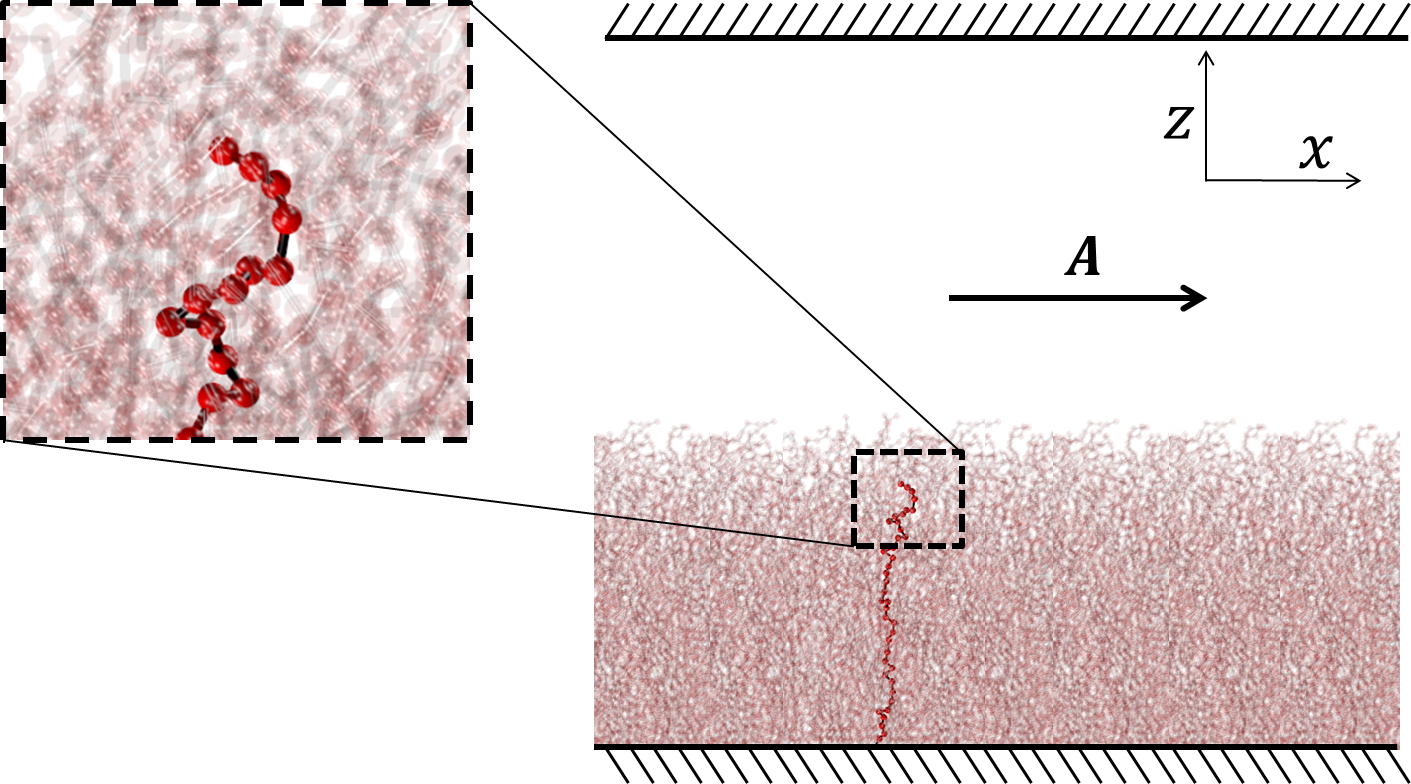}
\caption{Schematic representation of the investigated slit-pore system. Polymers are grafted at the bottom wall $z=0$ while the rest of the channel is occupied by solvent particles. An additional sketch is provided with a zoom on a typical chain (highlighted in red). The constant acceleration $A$ is applied along the $x$ axis to all solvent particles.}
\label{fig:system}
\end{figure}

A short description of the DPD methodology and of its application to the slit-pore geometry is
reported in the Supplementary Informations (S.I., Sect.1). Here we recall that  
we simulate a parallelepiped box of sides $L_x$, $L_y$, $L_z$, with polymers chains composed by $40$
%forty 
(coarse-grained) monomers grafted at $z=0$ (see Fig.~\ref{fig:system}).
%All the simulations are performed 
Our discussion pertains simulations in a box of size $L_x=30$, $L_y=5$ and $Lz=50$, in units of $r_c$ (see S. I. for definitions). 
We remark that we observe the same phenomenology we are going to present also in considerably bigger boxes ($L_x = 360$, $L_y = 20$ and $L_z = 150$). However, for the sake of computational time and memory capacity, the systematic analysis has been conducted in a smaller channel.
%In these units, 
The equilibrium distance between 
neighbour monomers in a chain is $0.89 r_c$. 
The grafting density  $\sigma_{graft}
\equiv N_{ch}/(L_xL_y)$ with $N_{ch}$ the number of chains composing the brush, 
being $\sigma_{graft}=1.5$, corresponds to the condition of dense brush.  Periodic boundary conditions
are applied along the $x$ and $y$ directions.  In the channel, a flow is imposed by applying a constant
acceleration $A\hat{x}$  to all solvent particles, resulting in a parabolic  velocity   profile along the $z$ direction.  
The strength of the velocity is  controlled by the value of $A$.  In the following, instead of $A$,
we will use the so-called  Weissenberg number $\displaystyle Wi \equiv  \frac{t_{brush}}{t_{flow}}$, which provides an adimensional information on the relative timescale of the 
%internal 
unperturbed brush dynamics 
($t_{brush}$) over the inverse of the averaged shear rate inside the channel $(t_{flow})$.    
Specific values for all parameters and their units and a detailed discussion of the $A$ dependence of ${t_{flow}}$
are reported in the S.I., Sect.1,3. 

%It should be noted that the mentioned experiments are performed in the region for which the flow intensity does not affect the brush height, therefore the two phenomena of backflow and of increased viscosity seem to be absolutely independent.
\begin{figure}[ht]
\centering%
%\subfigure[\label{fig:profile}]{
\includegraphics[width=8.cm]{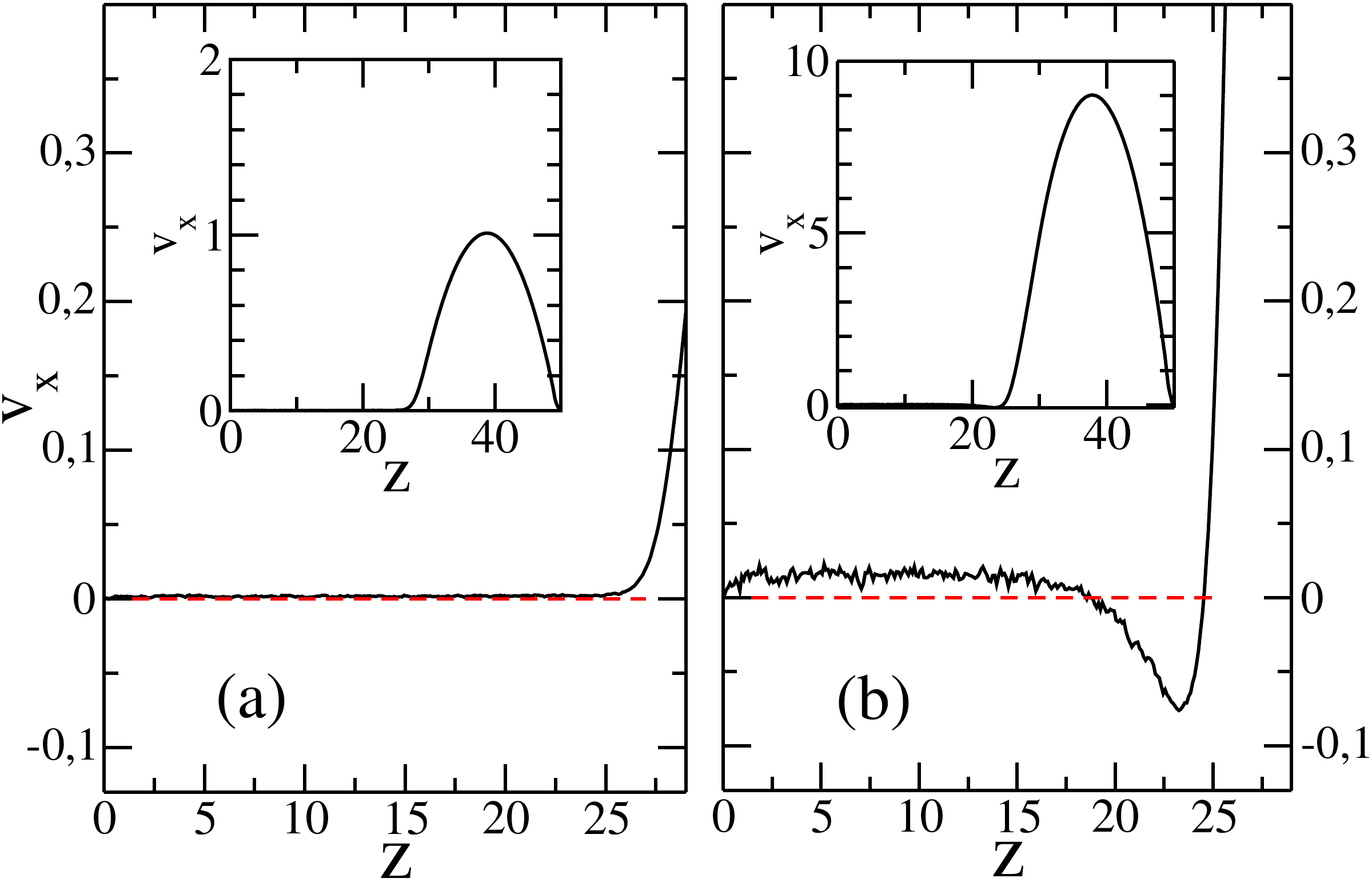}
%} 
\centering%
%\subfigure[\label{finestre}]{
\includegraphics[width=8cm,height=5cm,keepaspectratio]{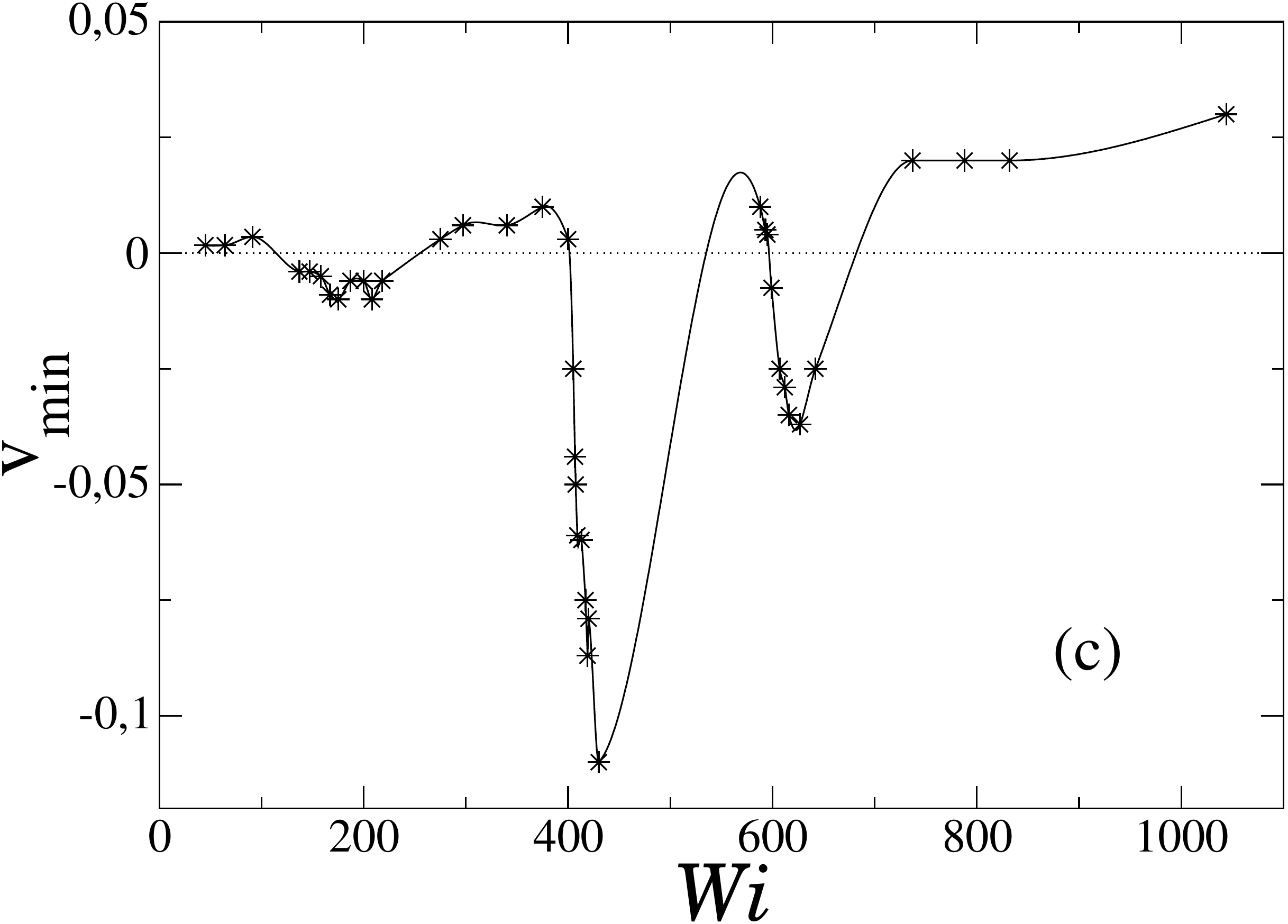}
%} 
\caption{%\baselineskip=12pt
Panels (a) and (b) show the velocity profile in the region where the brush is set and they are a zoom of the respective insets. In both cases, corresponding to (a) $Wi=45$ and (b) $Wi=418$, 
the profile is parabolic around $v_{max}$. In (a) no backflow is registered, while in (b) the flow inversion at $z \sim h$ is evident. Panel (c) reports the minimum value $v_{min}$ 
of the $v_x(z)$ profile for different $Wi$ (symbols). The line is a guide for the eye.
It is possible to recognize three regions for which $v_{min}$ assumes a negative value, signalling the presence of a backward flow.}
\label{fig:trend}
\end{figure}

The velocity profile $v_x(z)$,  calculated by averaging the $x$-component of the velocity for each solvent particle as a function of its position $z$, 
is shown in Figs.~\ref{fig:trend}.(a)-(b) for two  $Wi$ values, providing 
%respectively examples of absence or presence of flow inversion.   
two typical examples.
As expected, the functional shape of $v_x(z)$ in the region around the largest velocity $v_{max}$ is well represented by a parabolic function with the expected amplitude. 
In all cases,  the parabolic function predicts that  $v_x$ vanishes when $z$ approaches the brush height $h$,
suggesting that, for all studied $Wi$,  the presence of a dense polymer brush restricts the pore by an amount equivalent to $h$.
%~\cite{XXXX}.   
The velocity profile in the  region $z<h$, i.e. 
the region occupied by the brush,  is particularly interesting.  In some cases, e.g. the one of Fig.~\ref{fig:trend}.(a), the fluid velocity inside the brush is small, consistent with a picture in which the hydrodynamic interactions are effectively screened by the presence of the polymer layer.
In other cases,  as shown in Fig~\ref{fig:trend}.(b),   the velocity profile exhibits a flow inversion at the interface between the brush and the bulk, e.g. around $z \sim h$, 
as previously documented  for an analogous system~\cite{Deng} as well as when the solvent is replaced by a polymer melt in \cite{Leonforte,Muller_Pastorino}. In these previous studies,
the onset of flow inversion was associated to the point at which   the shear rate exceeds a certain threshold \cite{Deng}. 
Oppositely, we find that flow inversion takes place in different distinct disconnected windows of $Wi$ values. 
Indeed,  we uncover at least three different windows of flow intensities in which a backflow 
is observed. In between 
these regions the velocity profile at $z \sim h$ resembles the one in Fig.~\ref{fig:trend}.(a).  
Fig.~\ref{fig:trend}.(c) shows the minimum value $v_{min}$ of the profile $v_x(z)$ near the surface of the polymer brush, as function of $Wi$. By this definition, flow inversion 
is observed when $v_{min} < 0$.   The identified three back-flow regions fall around $Wi \sim 200,400,600$. 
 % We prefer to keep it well separated from the other ones since it presents different characteristics, as we will clarify later on. %Since it corresponds to very high flow intensity (see Sect.\ref{conclusion} for physical units), outside the physiological range of blood circulation to recall the leading example of endothelial glycocalyx, we rather focus on the first flow inversion and leave the second one to further studies not treated in this paper.
%We will focus on the first cases of inversion and leave the fourth one to further studies not treated in this paper.
%In all cases,  $|v_{min}|$ is of the order of $1\%$ of $v_{max}$.  Previously, it was argued~\cite{Leonforte,Muller_Pastorino,Deng} that backflow arises as a  consequence of the single chain dynamics.

In the previously mentioned papers \cite{Leonforte,Muller_Pastorino,Deng} it was argued that the backflow is a consequence of the grafted single chain dynamics, performing  a recursive  imperfect cycle composed of: tilting, 
elongation  and recoiling.  Indeed, an elongated chain (and thus exposed to flow) is dragged by the shear stress along the flow direction and then recoiled back by entropic forces.  By examining the polymer trajectories we confirm
the presence of such recursive motion (see movie in the S.I., Sect.4) for all  $Wi$  values,
but such cyclic motion is found to be  independent on the presence of the back-flow effect.  Hence, such single-polymer motion can not explain the
onset of flow inversion. 

%\footnote{We reject the use of \textquotedblleft tumbling\textquotedblright and of \textquotedblleft cyclic motion\textquotedblright, used in literature, 
%Deng lo usa citando muller_pastorino, lo usa muller_pastorino e leonforte! for the following reasons: \textquotedblleft tumbling\textquotedblright is usually meant to indicate the rolling of a body immersed in a shear flow~\cite{tumblingRBC_chaouqi,tumbling_polymer}, while the brush is for the most of its height screened to hydrodynamics; \textquotedblleft cyclic motion\textquotedblright is properly associated to a periodic motion, while the brush cannot recover perfectly previous configurations.}

\begin{figure}[ht]
\centering%
\subfigure[\label{snapshot1}]{
\includegraphics[width=3.7cm,height=1.5cm]{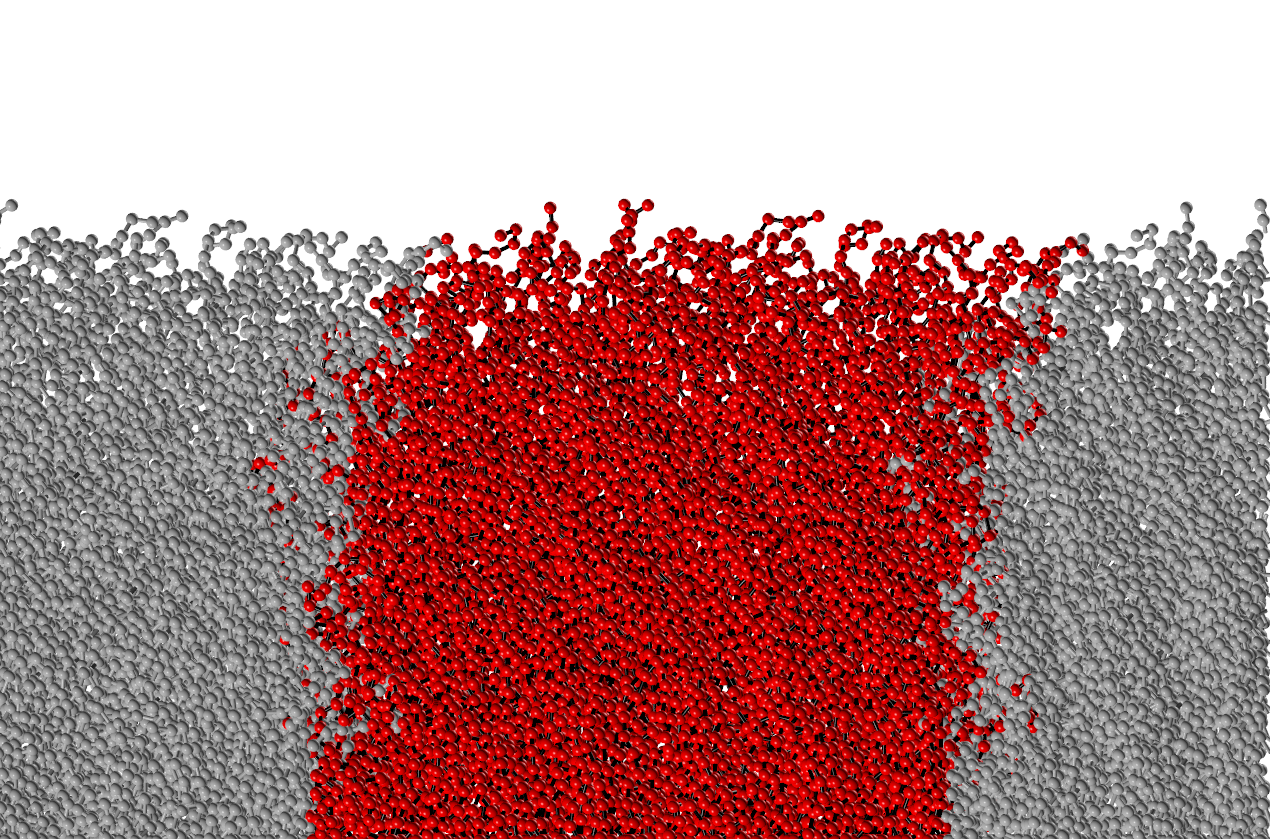}
} 
\subfigure[\label{snapshot2}]{
\includegraphics[width=3.7cm,height=1.3cm]{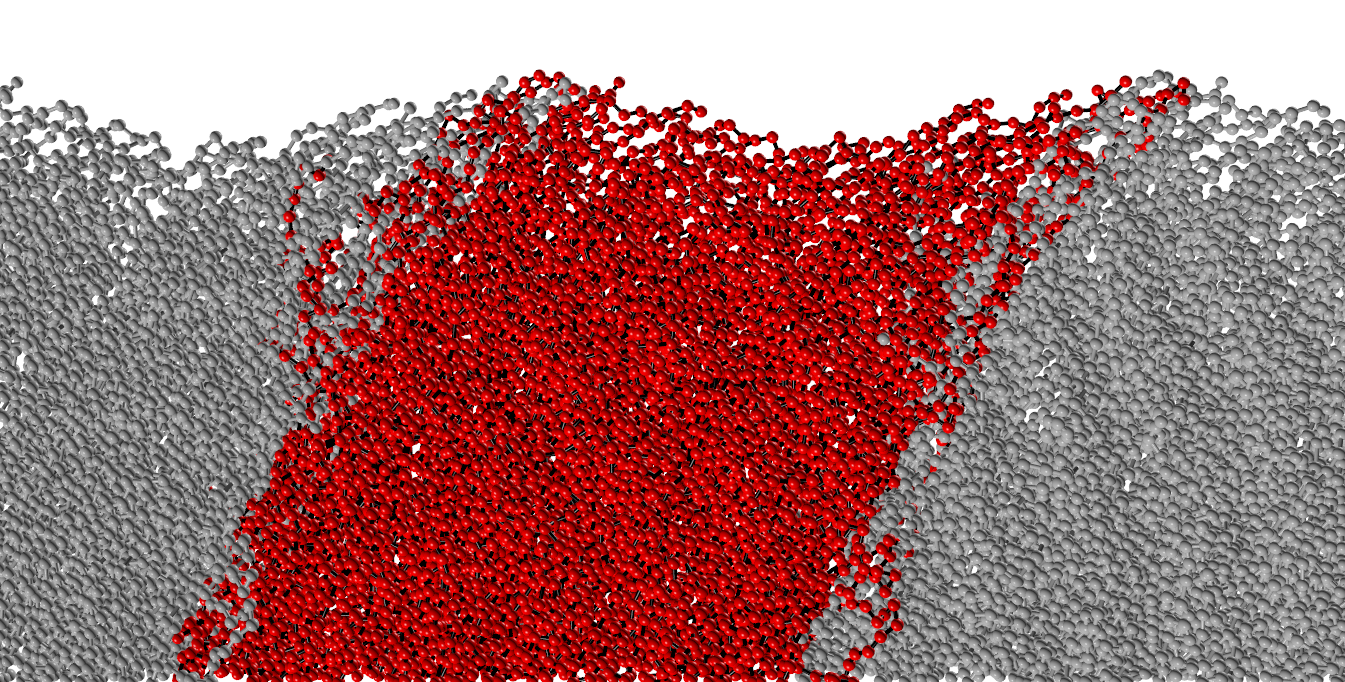}
} \\ 
\subfigure[\label{snapshot3}]{
\includegraphics[width=8cm]{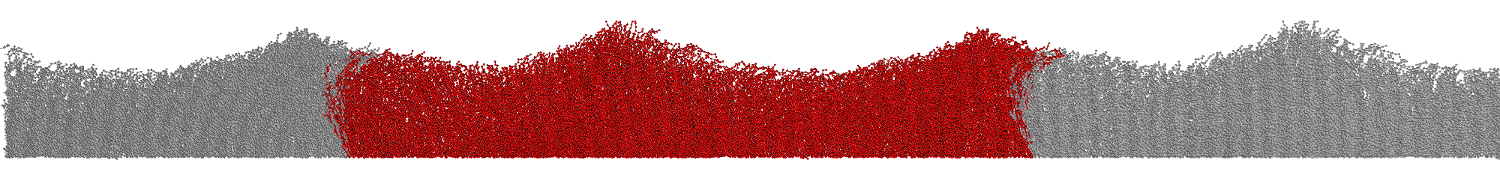}
} 
\caption{%\baselineskip=12pt
Sketches of the whole brush, corresponding to (a) $Wi=91$ and (b) $Wi=418$. In (a) the brush surface is basically flat, while in (b) the surface is modulated by a traveling wave with $\lambda \sim L_x=30$, $\nu=0.077$ and $b=1.18$. In (c) a snapshot for a larger studied systems ($L_x=180$, $L_y=5$ and $L_z=150$) with a surface wave with multiple crests of wavelength $L_x/2 \approx 90$ ($Wi=340$). We show the central box (in red) and the two nearest replicas (in grey) to highlight the periodic nature of the waves.}
\label{snapshots}
\end{figure}

%\begin{figure}[ht]
%\centering%
%\includegraphics[width=8cm]{sketch_onde_cut.png}
%\caption{
%\green{Simulation snapshots for the largest studied systems ($L_x=180$, $L_y=20$ and $L_z=150$). We show the central box (in red) and part of the two nearest replicas (in grey).
% for the two topmost panels. 
%(Top) No surface waves ($Wi=3$)
%and (bottom)
%%, (middle) 
%a surface wave with a wavelength of $L_x/2 \approx 90$ ($Wi=340$).
%% and (bottom) a surface wave appearing for the same system, with the difference that polymeric chains do not experience periodic boundary conditions along $x$.
%}
%}
%\label{snapshots}
%\end{figure}

\begin{figure}[ht]
\centering%
%\subfigure[\label{frequency}]{
%\includegraphics[width=4cm]{INSETonda005.pdf}}
%\subfigure[\label{wavelength}]{
%\includegraphics[width=4cm]{INSET_lambda005.pdf}
\includegraphics[width=8cm]{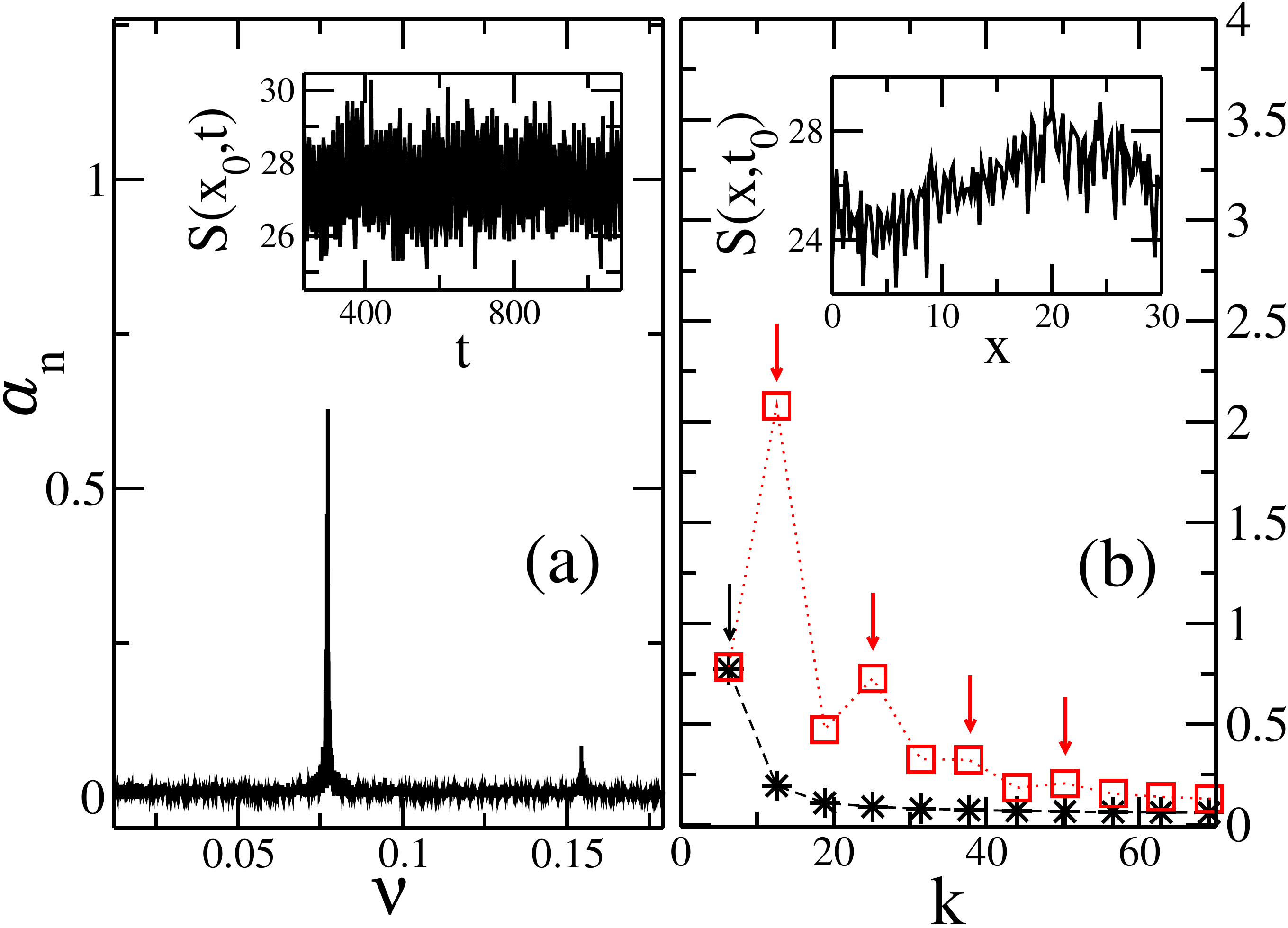}
%} 
\centering%
%\subfigure[\label{vu_vs_Wi}]{
\includegraphics[width=8cm,height=5cm,keepaspectratio]{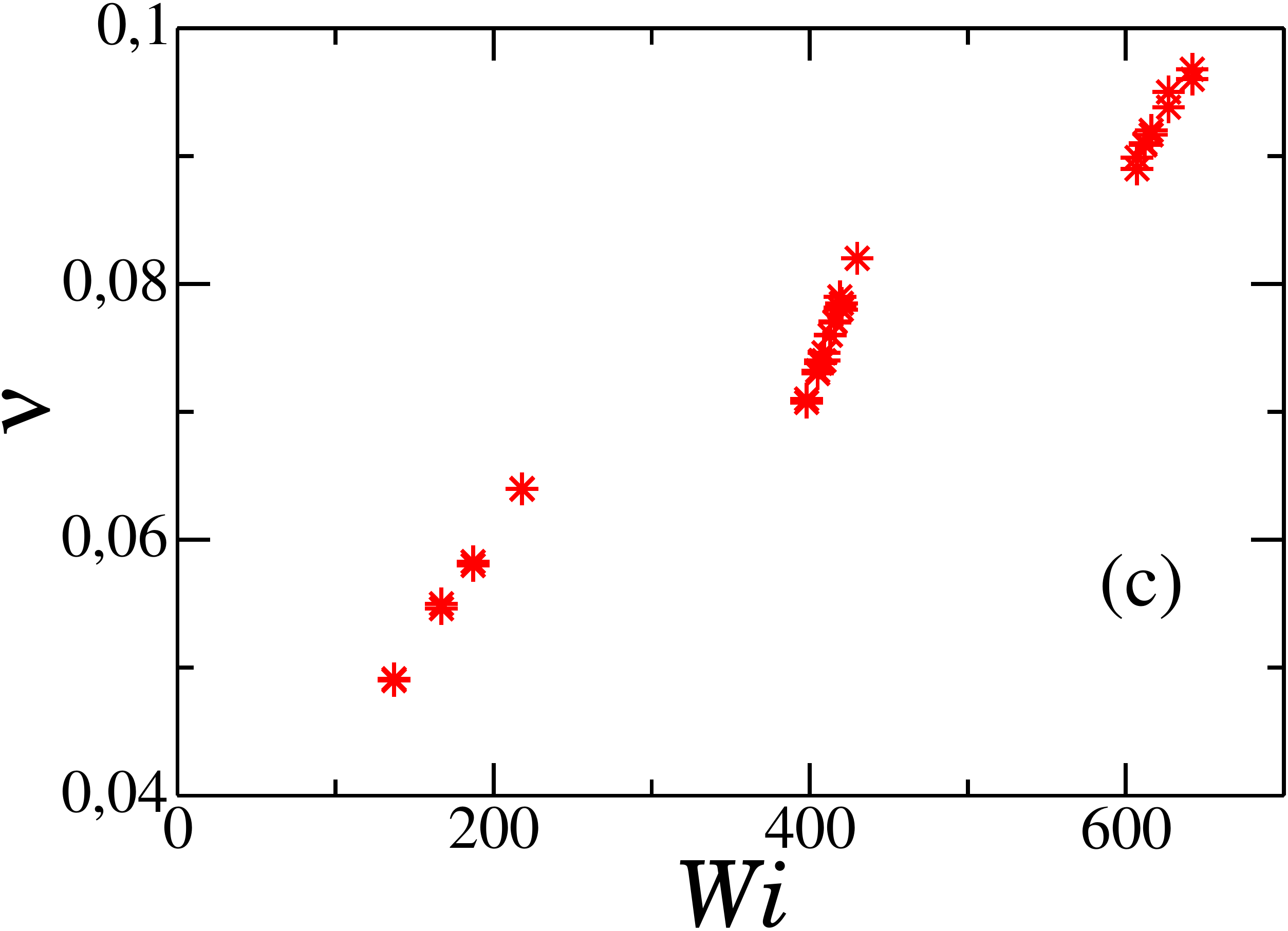}%} 
\caption{%\baselineskip=12pt
Fourier power-spectra: (a) in the frequency domain $\nu$ for the signal $S(x_0,t)$, 
(b) in the wavenumber domain $k=2\pi/\lambda$ for the signal $S(x,t_0)$ (red squares indicate the spectrum in a bigger system $L_x=180$, $L_y=5$ and $L_z=150$). In both cases the first and second harmonics are visible.
The two insets show typical signals in real space. %(in (b) the faint peak around $k \approx 7.5$ represents the local structure of the brush, namely the average distance between chains $\lambda \sim d_{graft} \approx 0.82$). 
Panel (c) shows the first harmonic frequencies corresponding to the $Wi$ giving backflow. Such frequencies monotonically increase.}
\label{fig:wave_features}
\end{figure}

Investigating space and time correlations among different polymers, we discover a wave traveling over the brush surface in the same direction as the imposed flow. The wave arises {\it only} when back-flow is present, suggesting a strong association between  the surface wave and the capability of the brush to produce an inversion of the flow velocity around $z \sim h$.
A visual observation of the brush and of its dynamics nicely show the collective behavior of the polymers:
Fig.~\ref{snapshots}.(a),(b)  show a typical
frame for  two generic cases of absence and presence of flow inversion.
Videos of the time dependence of the brush are available in the S.I., Sect.4. We find that this traveling wave has a clear spatial periodicity
of the order of the simulation box and a non-negligible amplitude.
We characterize  such surface wave evaluating its frequency $\nu$, wavelength $\lambda$ and oscillation amplitude $b$. To do so we define 
 the brush surface $S(x,t)$ 
 as the position of the furthest monomer from the grafting wall occupying at time $t$ the coordinate $x$, averaged over all $y$.   The power-spectrum of the time Fourier transform of $S(x_0, t)$  (for an arbitrary $x_0$ value, inset of 
 Fig.~\ref{fig:wave_features}.(a) )
  displays, in all cases in which a flow inversion is observed,  a sharp peak at a given frequency (Fig.~\ref{fig:wave_features}.(a)) that we associate to $\nu$.  In  some cases a much less intense peak is also
  observed at $2 \nu$.     
 Fourier transforming the signal $S(x,t_0)$ in real space for an arbitrary $t_0$ and averaging over all configurations
 provides a quantification of the wavelength $\lambda$.  As shown in Fig.~\ref{fig:wave_features}.(b), for all cases in which a flow inversion is present we find, as in the time domain, 
 a dominant contribution from the
longest $\lambda$ that can propagate in  a period system of size $L_x$ and from its second harmonic 
(e.g. from wavenumbers $k$ equal to $\frac{2\pi}{L_x}$ and $\frac{4\pi}{L_x}$).   
 Fig.~\ref{fig:wave_features}.(c) shows  the $Wi$ dependence of $\nu$ for systems with the same $L_x$ (e.g. same $\lambda$).
 In each of the three disconnected regions of $Wi$ values where flow inversion is present, $\nu$ increases
 approximatively linearly with $Wi$.    Finally, we define the  wave amplitude $b$  as the standard deviation of  $S(x,t)$
 and the wave average height as $\bar S  \equiv < S(x,t)>$. 

We stress
%have checked 
that wave propagation is observed also in systems of different  $L_x$. 
%Indeed, 
We find that increasing the length $L_x$ and the width $L_z$ of the channel, multiple wave crests can be displaied (in Fig.~\ref{snapshots}.(c) an example).
%the propagation 
%speed $U$, defined as the product $\lambda \nu$, as well as the wave amplitude remain constant. By contrast, $v_{min}$ decreases on increasing $L_x$ (therefore increasing $\lambda$).
This suggests that the finite size of the simulation box 
exerts a cutting on the density of states,
%cuts the density of states at the frequency corresponding to $\lambda \approx L_x$, 
making it possible to observe only waves with wavenumbers 
which are integer multiple of $2\pi/L_x$. Further details on finite size effects are reported in the S.I., Sect.5. 

%The two natural questions are: i) Hhow can the wave form? ii) Why is the wave in concurrence with the backflow?
%We propose our answers in the following paragraphs.

%that the wave properties are consistent with predictions derived by Taylor in his seminal study on pusher microswimmers.

%To explain such an association between back-flow and wave we propose to consider the brush as an anchored microswimmer, unable to move but able to propel the surrounding liquid. Literature explains that ciliated micro-organisms, when their cilia act together in metachronal synchronization~\cite{Gompper2013,Blake1972,alga_volvox}, develop a self-propelling ability. 
%However, as explained in Sect.\ref{wave_formation}, active cilia are not the best reference model for our system.  We prefer to treat the brush surface as a continuum, thus a more suitable analogy is provided by a Taylor's study~\cite{Taylor},

\begin{figure}[ht]
{
\includegraphics[width=8cm,height=5cm,keepaspectratio]{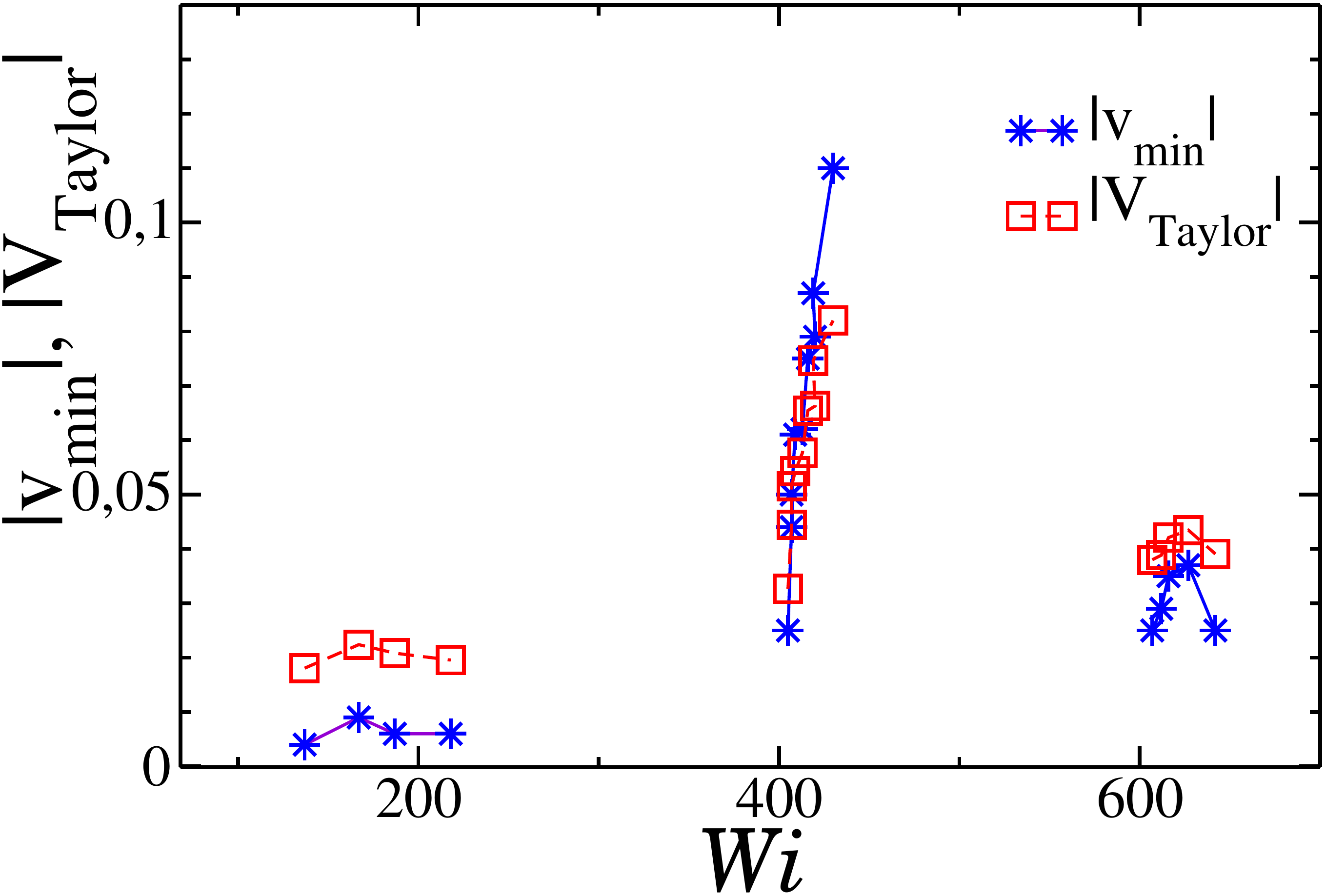}}
\caption{%\baselineskip=12pt
The mapping between the absolute value of the velocity profile minimum $|v_{min}|$ (stars) and the velocity $V_{Taylor}$ obtained by inserting in Eq.~(\ref{eq:propulsion}) 
the wave amplitude $b$, the wavelength $\lambda$ and the speed velocity $U=\lambda\nu$ evaluated for the brush surface wave at the given $Wi$.
%self-propelling velocity of a Taylor's microswimmer (squares) actively producing a wave.
}
\label{taylor}
\end{figure}

To dig into the reasons why a flow inversion is observed every time there is a wave propagation,
we recall the derivation provided  by Taylor~\cite{Taylor}, dated back to 1951,  to explain the mobility of pusher microswimmers. 
In his seminal paper, Taylor determines analytically the self-propelling velocity $V_{Taylor}$, at low Reynolds number, of an infinite two dimensional sheet 
whose profile is a sinusoidal wave traveling with
%(inverted)
(opposite) velocity $U$:
\be
\label{eq:propulsion}
|V_{Taylor}|=\frac{2 \pi^2 b^2 }{\lambda^2}\left( 1-\frac{19}{4}\frac{\pi^2 b^2}{\lambda^2} \right) U ,
\ee
%problemi con i segni!!!!!!!!!!!!!!!!!!!
where $b$ and $\lambda$ are, respectively, the amplitude and wavelength of the sinusoidal wave.  
By associating the  two dimensional sheet with the brush surface, the
wave velocity of the sheet with the wave velocity of the brush surface
and the self-propelling velocity of the
sheet with the 
%(inverted) 
opposite velocity of the solvent at the brush surface $v_{min}$, 
Taylor's expression provides a precise relation between the parameters entering in the 
wave phenomenon and the parameters controlling the flow inversion.  
In Fig.~\ref{taylor} we show, for each set of $b$, $\lambda$ and $U$ values associated to a specific $Wi$ value
giving flow inversion, both $|V_{Taylor}|$ (from Eq.~(\ref{eq:propulsion})) and $|v_{min}|$.  All the trends of the $|v_{min}|$ behavior are perfectly recovered by the self-propelling velocity of a Taylor's micro-swimmer, 
suggesting that
indeed the elastic brush can be  considered as an anchored micro-swimmer, unable to move but able to propel the surrounding liquid.

\begin{figure}[ht]
{
\includegraphics[width=7cm]{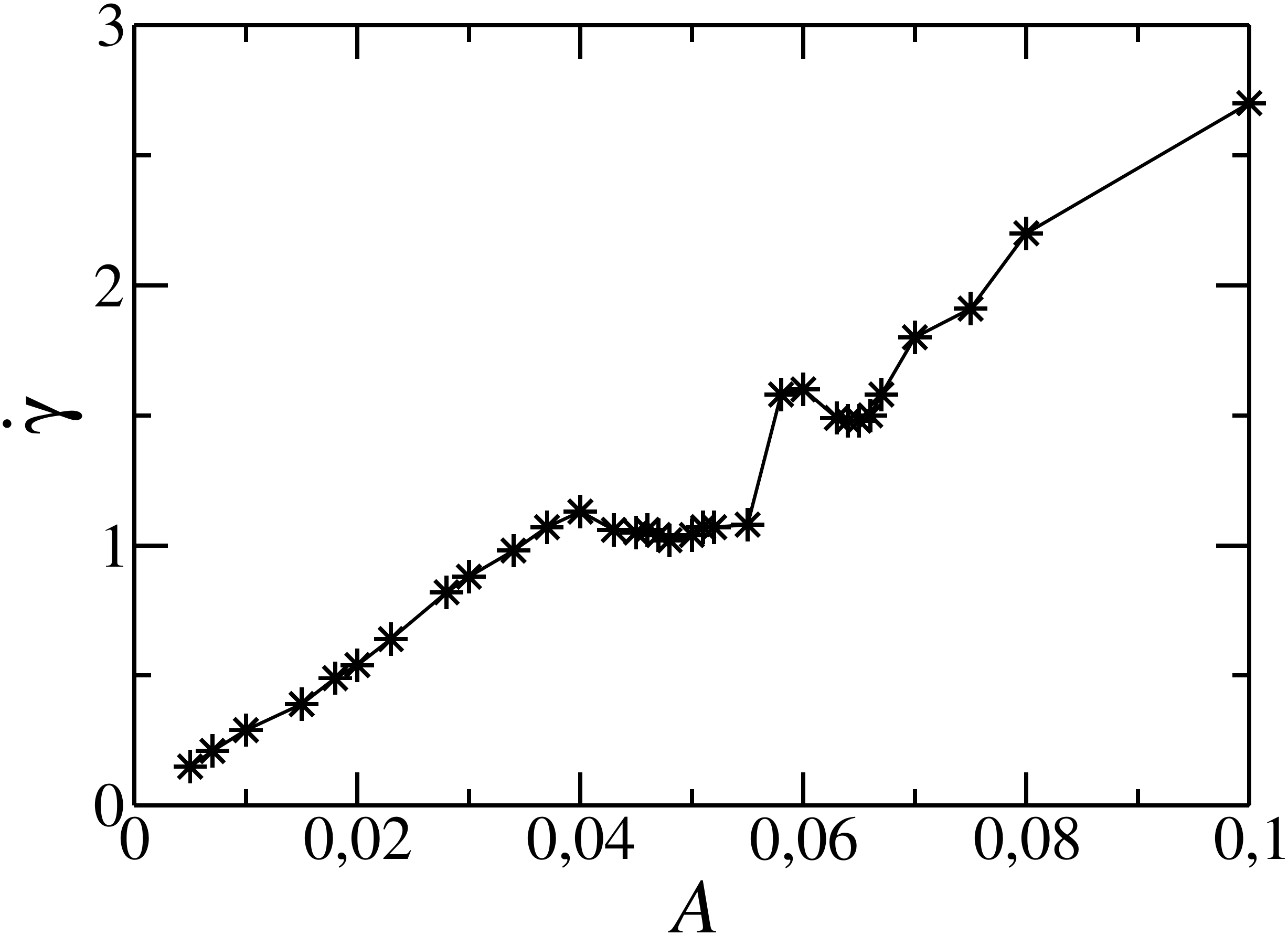}}
\caption{%\baselineskip=12pt
The shear rate $\dot{\gamma}$ measured from a linear fit of the velocity profile $v_x(z)$ at the brush surface $z \sim S$ for different flow intensities $A$.}
\label{elasticity}
\end{figure}

As a last consideration, we investigate possible origins of the surface wave. 
Our results suggest that the wave propagation arises from 
hydrodynamics interactions synchronising polymers, in a sort of metachronal motion.
A \textquotedblleft metachronal wave\textquotedblright is known to develop in ciliated systems, namely flat surfaces covered by a regular matrix of equally 
spaced flexible filaments~\cite{Gompper2013,Blake1972}. Exposed to flow, filaments do not interact directly, but thanks 
to hydrodynamics interactions, after some transient time, they reach a synchronized state with a regular phase shift between subsequent filament rows. 
However, in those cases, the matrix is composed by active matter elements and the sequence of configurations they assume is assigned \textit{a priori}. 
On the contrary, polymer brushes are passive media. Different approaches have been developed to model wave formation on 
%elastic 
passive media such as 
crop canopies (because of flowing wind) or aquatic vegetations (because of water flow), but they require also the inertial 
term of Navier-Stokes equations to account for the instability initiating the surface modulation~\cite{gosselin_delangre_grano,flow_over_elasticsurfaces}.
However, we can consider the brush to be an elastic medium. Therefore, the presence of a wave suggests that a possible
resonance effect may arise between the shear rate $\dot{\gamma}$ at the brush surface, produced by the imposed flow, and the frequency $\nu_{com}$ associated to the 
time scale of relaxation from the elastic compression of the whole brush~\cite{mahadevan_flutter}.  
This would point to one threshold
%specific 
value of $Wi$ beyond which surface oscillations are observed,  which is not
consistent with the finding of waves in at least three different $Wi$ regions.
To clarify this inconsistency,  we show in Fig.~\ref{elasticity} 
the shear rate $\dot{\gamma}$ estimated  by a linear fit of the velocity profile $v_x(z)$ at the brush surface $z \sim S$, 
as a function of $A$.  
We ascribe the striking non-linear behaviour of $\dot{\gamma}$ vs. $A$ to the compression of the brush at high imposed flow.
We also find a significant dependence of the brush elastic compression properties on $A$, suggesting the possibility of multiple intersections of 
the two quantities $\dot{\gamma}$ and $\nu_{com}$ (and hence multiple resonances and multiple
regions of flow inversion) on varying $A$ or, equivalently, $Wi$ (see S.I., Sect.2).

In summary, 
%\sout{we have studied the behaviour of dense polymer brushes under parabolic flow by means of numerical simulations. Increasing $Wi$ we have detected three ranges of imposed flow for which a  velocity inversion appears at the interface between the brush and the rest of the channel.} 
we have discovered for dense polymer brushes under flow a well-characterized surface traveling wave, 
which appears  in all cases of back-flow and it is not present in the other ones, providing a
novel picture of the flow-inversion phenomenon, associated to a  collective motion of the whole brush rather than to
a single chain dynamics. We have also observed a striking and unexpected similarity in the relation between the wave 
velocity and the back-flow velocity in the case of flow in passive polymer brushes and the same relation for the 
%classical 
case
of active pusher microswimmers~\cite{Taylor}.

%\begin{acknowledgments}
Discussions with prof. Philippe Peyla are kindly acknowledged. C.M. and S.B. also acknowledge financial support from CNES (Centre National d'Etudes Spatiales) and ESA (European Space Agency). 
L.R. acknowledges support from the Austrian Research Fund (FWF) through his
Lise-Meitner Fellowship M 1650-N27.
%\end{acknowledgments}

%\bibliography{biblio_wave}

\clearpage
\begin{center}
\begin{LARGE}
{\bf Supplementary Information}
\end{LARGE}
\end{center}

\vspace{0.8cm}

\section{1. DPD methodology}
DPD is a coarse-grained Molecular Dynamics method. It was introduced by Hoogerbrugge and Koelman in 1992~\cite{Hoogerbrugge_Koelman} to simulate isothermal Navier-Stokes equations and it has been recently recognized that it can be successfully applied to any mesoscopic scale~\cite{Nm_for_DPD}. In the spirit of DPD, the simulated particle does not correspond to a single molecule, but rather to a significant large  \textit{cluster} of them.   
Each of the $N$ point-like particles evolves in time according to 
Newton's equations 
\be
\label{Newton}
m_i \dot{\vec{v}}_i=\sum_{j \neq i }^N \vec{F}_{ij}=\sum_{j \neq i }^N (\vec{F}_{ij}^C+\vec{F}_{ij}^D+\vec{F}_{ij}^S)
\ee
that we solve by using the velocity Verlet algorithm~\cite{Frenkel}.  
Observables are then calculated as averaged all over time configurations. 
The force $\vec{F}_{ij}$ of Eq.~\ref{Newton} on each particle has three contributions~\cite{bridging_the_gap} resulting from the coarse-graining procedure: a \textit{conservative} component $\vec{F}_{ij}^C$, a \textit{dissipative} component $\vec{F}_{ij}^D$ and a \textit{stochastic} one $\vec{F}_{ij}^S$.
All these forces are pair-wise, to guarantee momentum conservation, and limited to a cut-off radius $r_c$.

The conservative  soft-core repulsive  force has the following expression:
\be \label{forza_cons}
\vec{F}^C_{ij}= 
\begin{cases} 
a_{\alpha,\beta} \left( 1- \frac{r_{ij}}{r_c} \right) \hat{r}_{ij}  &   r_{ij} \leq r_c , \\ 
0    &  r_{ij} > r_c 
\end{cases}
\ee
with  $\vec{r}_{ij}=\vec{r}_i - \vec{r}_j$ is the vector distance between the i-th and j-th particle, $r_{ij}=|\vec{r}_{ij}|$ and $\hat{r}_{ij} =\vec{r}_{ij}/r_{ij}$.
%The choice for the soft-core potential is justified by the \textquotedblleft blob\textquotedblright nature of DPD particles that, containing (in the atomistic perspective) also empty space, should have the possibility to overlap; the choice for the repulsive effect has basis in liquid theories, stating that the specific structure of a liquid is fundamentally determined by the repulsive part of the intermolecular potential (while the attractive part can be treated as perturbation)~\cite{Forrest_Suter}: 
Such repulsive potential  models the averaged fast microscopic length and time scales and was first derived
by coarse-graining particles interacting via a  Lennard-Jones potential~\cite{Forrest_Suter}.
%(it is assumed that microscopic scales are well separated (fast) and indipendent from the other ones (slow)).
The indexes ${\alpha,\beta}$ indicates the particle type (solvent or polymer, in our case). 
The constant $a_{\alpha,\beta}$ measures  the force between two completely overlapping particles and it is casted from the compressibility of the modeled fluid.
%\footnote{DPD in its common formulation is a top-down coarse-graining approach. However, consecutively, it has been justified also by means of bottom-up derivations~\cite{Espanol2001,Espanol_hydro,bottom-up,Kinjo_microDPD,fokker-planck}.}
%NB: top-down perché non solo si derivano i valori dei parametri da grandezze macro, come la compressibilità, ma anche perché i fattori di scala sono calibrati in modod da dare le giuste viscosità e diffusività!

The two other forces account for the loss of details in the coarse-graining procedure. 
% and attempt to reproduce   diffusivity and viscosity of the modeled fluid.
%, namely the huge number of collisions occurring between real  molecules and constituting the microscopic foundation of transport properties, such as viscosity and diffusivity
%\footnote{nota sullo Schmidt number? basta cmq un approccio top-down per le scale fisiche... Thermodynamics of a DPD system resemble a gas. The Schmidt number, defined as the ratio between momentum diffusivity over mass diffusivity, is order 1 instead of order $10^3$, as it should be for a liquid.}.
% bella spiegazione del numero di Schmidt = 1 in Groot_Rabone, pag 728!!!
The dissipative force has the form:
\be \label{forza_diss}
\vec{F}^D_{ij}= -\gamma w^D(r_{ij}) (\hat{r}_{ij} \cdot \vec{v}_{ij}) \hat{r}_{ij} ,
\ee
where the standard choice in literature for the \textquotedblleft weight function\textquotedblright $w^D(r_{ij})$ is: 
\be \label{peso_diss}
w^D(r_{ij})= 
\begin{cases} 
\left( 1- \frac{r_{ij}}{r_c} \right) \hat{r}_{ij}  &   r_{ij} \leq r_c , \\ 
0    &  r_{ij} > r_c .
\end{cases}
\ee
Equation (\ref{forza_diss})  introduces a friction among particles proportional to the relative velocity $\vec{v}_{ij}=\vec{v}_i - \vec{v}_j$ 
%and to a factor $\gamma$ that will depend on the temperature $T$. 
Thermal fluctuations are added via the stochastic force: 
\be
\label{forza_stoc}
\vec{F}^S_{ij}= \sigma w^S(r_{ij}) \theta_{ij} (\Delta t)^{-\frac{1}{2}} \hat{r}_{ij}
\ee
where $\sigma$ is a constant (related to temperature), $w^S(r_{ij})$ is a weight function and
%\footnote{$\sqrt{\Delta t}$ due to the discretization of a Wiener process.}
$\theta_{ij}$ is a 
random number extracted from a gaussian distribution with  zero average and 
 indipendent in time and among particle pairs: $\langle \theta_{ij}(t) \theta_{lm}(t') \rangle = (\delta_{il} \delta_{jm} + \delta_{im} \delta_{jl})\delta(t-t')$. 
Momentum conservation requires  $\theta_{ij}=\theta_{ji}$.

Two additional constraints, $w^D(r_{ij})=[w^R(r_{ij})]^2$ and $\displaystyle \gamma=\frac{\sigma^2}{2k_BT}$, guarantee that the probability distribution respects the statistics of the NVT ensamble~\cite{Espanol_Warren}.
%\footnote{To be noted that $\vec{F}^D(r)$ and $\vec{F}^S(r)$ can be thought of as a thermostat assuring isothermal hydrodynamics.} 
%WAC THEORY
%Thermodinamic quantities are calculated by means of this conservative force.
%The other two forces, in fact, can be tought of as a thermostat, guaranteeing a constant temperature, once the stationary state has been reached.
%The viscous force tends to reduce particle relative velocity (it cools down the system), while the random force synthesizes collision (it warms up the system).
%The net force (risultante?) is not conservative but central, therefore energy is not conserved while momentum is.

To mimic the brush chains we add a 
finite extensible nonlinear elastic potential (FENE)~\cite{why_fene} 
\be
\vec{F}^{fene}_{ij}= -2k R^2 \frac{r_{ij}-r_{eq}}{R^2-(r_{ij}-r_{eq})^2} \hat{r}_{ij}~~~~~~ r_{ij}-r_{eq}<R
\label{fene}
\ee
which account for the neighbour particle connectivity. In Eq.~(\ref{fene}), 
$r_{eq}$ is the equilibrium distance between neighbour monomers and $R$ is the maximum allowed extensibility. Grafted points are randomly chosen from a uniform distribution and located on a flat surface. Note that polymers are 
 non-ideal since they interact via an excluded volume potential. %they are not completely flexible

\subsection{1.1 DPD and physical units}
We chose the DPD units such as $r_c=1$, $m_i=1$, $t_{DPD}=1$ and $k_BT=1$. %With these definitions the number of free parameters are .......\ref{Espanol2001}.
Following~\cite{bridging_the_gap}, we fix the solvent-solvent interaction parameter $a_{SS}$  relating it to the  adimensional compressibility of water $\kappa_T^{-1}$.  A comparison between the equation of state of a simulated DPD system and the experimental data  ($\kappa_T^{-1}(water)=16$ at room temperature) suggests $a_{SS}=\displaystyle 75 {k_BT}/{\rho}=25$.  We assume that the polymer-polymer  interaction parameter has the same value (e.g. $a_{SS}=a_{PP}$), while we select a  smaller value for the solvent-polymer parameter $a_{SP}=20$
(\textit{good solvent} conditions).  The noise amplitude is fixed to $\sigma=3$.
For the FENE potential, finally, we use $r_{eq}=0.86$, $R=1$ and $k=50$.

In DPD the choice for the physical units depends on the level of coarse-graining is desired. 
Our reference system is the endothelial glycocalyx, therefore we fix the physical lengthscale $l_{phys}$ relating the spacing between different filaments of the glycocalyx network, $d_{glyco} = 20nm$~\cite{regalino} with the average distance between anchor points of our brush, $d_{graft}=\sqrt{1/\sigma_{graft}}= 0.82$:
%the persistence length of one of the biopolymers present in the endothelial glycocalyx network to the equilibrium distance set in the FENE potential. For example, the hyaluronic acid has a persistence length around $30nm$, 
\be
l_{phys}= d_{glyco}/d_{graft}= 24 \cdot 10^{-9}m.
\nonumber
\ee
We underline that with this choice also the brush thickness is in the range of endothelial glycocalyx ($100nm$ \textdiv $1000nm$ \cite{EGL_thickness}).

As for the physical mass scale $m_{phys}$ and physical time scale $t_{phys}$, we exploit the comparison between viscosities and between energies. The viscosity of water $\eta_{phys}$ is $\eta_{phys}=10^{-3}Pa \cdot s$ at $300K$ and the DPD viscosity $\eta_{DPD}$ in a bare channel, estimated from the slit-pore velocity profile relation $\eta=\rho A (L_z/2)^2/2v_{max}$, where $v_{max}$ is the maximum velocity (see Sect.~1.2 for definitions of the other parameters), corresponds in our case to $\eta_{DPD} =0.84$.
The physical energy scale is defined as $k_B T_{phys}$, with $T_{phys}=300K$.
Thus we can write down two relations
\bse
\be
m_{phys}=\frac{\eta_{phys}}{\eta_{DPD}} t_{phys} l_{phys} ,
\nonumber
\ee
\be
t_{phys}=\frac{l_{phys}}{v_{phys}}=\frac{l_{phys}}{\sqrt{\frac{3 k_B T_{phys}}{m_{phys}}}} ,
\nonumber 
\ee
\ese 
from which we extract
\bse
\be
t_{phys}=\frac{l_{phys}^3}{3 k_B T_{phys}} \frac{\eta_{phys}}{\eta_{DPD}}=1.8 \cdot 10^{-6} s,
\nonumber
\ee
\be
m_{phys}=\frac{\eta_{phys}}{\eta_{DPD}} t_{phys} l_{phys}=5.1 \cdot 10^{-17} Kg .
\nonumber 
\ee
\ese

\subsection{1.2 The system geometry}
We simulate a parallelepiped box of sides 
$L_x=30$, $L_y=5$ and $L_z=50$.  On the $x$- and $y$- axes we impose periodic boundary conditions, while at $z=0$ and $z=L_z$ two impenetrable parallel walls of infinite mass are set. At the wall we use the so-called \textquotedblleft bounce back reflection\textquotedblright conditions~\cite{Revenga:boundary_in_dpd} in which all the velocity components are reversed:
\be 
\vec{v} \rightarrow -\vec{v} \quad \hbox{at} \quad z=0,L_z .
\ee
Such conditions assure no-slip boundaries. 
The integration  time step  is $\Delta t = 0.02$. We set  the density $\rho=\displaystyle N/(L_xL_yL_z)=3$~\cite{bridging_the_gap}, therefore  $N=22500$. 
We fix the number of monomers per chain $n=40$ and the grafting density $\sigma_{graft}=1.5$, defined as $\sigma_{graft}=N_{ch}/(L_xL_y)$, with $N_{ch}$ the number of chains composing the brush.
%The specific ratio between fluid particles and monomers will vary according to the grafting density and the chain length.
For  the sake of ease and computational time, we  attach polymers only at the bottom wall.
%, paying attention during data analysis in correcting adequately the loss of symmetry along the $z$-axis. 

In order to produce  a parabolic velocity profile  inside the channel  (e.g. along the $z$ direction) 
a constant acceleration $\vec{A}=A \hat x $ is applied to all fluid particles:
\be
m_i \dot{\vec{v}}_i^{fluid}=\vec{F}_{ij}^{fluid}+m_i \vec{A} .
\ee
In turn, these fluid particle exchange moment with the chains, dragging them.
Different values of $A$ allow us to probe different dynamic regimes. 

\section{2. Brush equilibrium properties}

\begin{figure}[ht]
\centering%
\includegraphics[width=8cm]{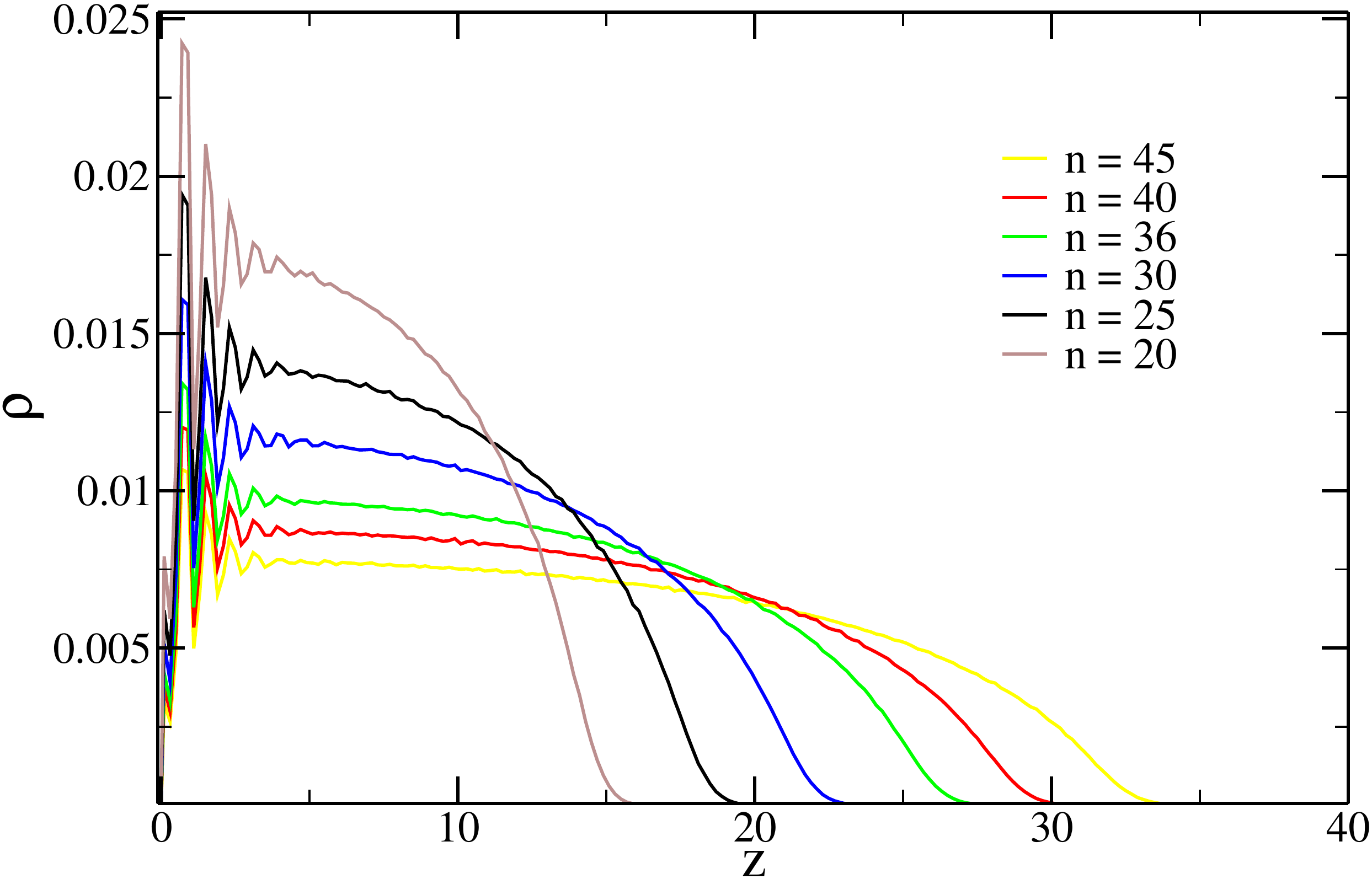}
\caption{
Probability distributions of polymer brushes at equilibrium: fixed the grafting density, $\sigma_{graft} = 1.5$, we varied the degree of polymerization $n$.
The case $n=40$ corresponds to an almost step-like profile $\rho(z)$.}
\label{density_prf_graft1.5}
\end{figure}

Polymer brushes immersed in a liquid at equilibrium (i.e. in the absence of flow) have been  thoroughly  investigated in the past
~\cite{Milner1988,Milner1991,Binder2011,barrat}.
Here we recall that the brush equilibrium conformation results from the balance between configurational entropy, that tends to make chains visit the whole available space, and excluded volume interactions, avoiding 
contact between monomers.  The brush can be properly described by its  profile $\rho(z)$, indicating
the probability distributions of finding a monomer at distance $z$ from the grafting wall.  
We show in Fig.~\ref{density_prf_graft1.5}   $\rho(z)$  at fixed $\sigma_{graft}=1.5$
for several chain lengths ($n=20,25,30,36,40,45$). 
In line with previous theoretical results~\cite{Milner1991} and numerical studies~\cite{Binder2011,Pastorino_Binder_2006},  
the higher  the volume fraction, the more step-like  the density profile is. 

A definition of the brush height $h$ in term of the first moment $<z>$ of the density profile depends on the shape of   $\rho(z)$.  Assuming that each chain is completely elongated so that free ends are all at the  maximum distance from the grafting wall, the profile $\rho(z)$ is step-like (\textit{Alexander model}) and the brush height can be defined as $h=2<z>$. A less rigid theory (\textit{self-consistent field theory}) hypothesizes that monomers of a same chain are distributed as a random walk, therefore the free end position ranges uniformely over the whole brush thickness. In such a case the profile is parabolic, $\rho(z)=\rho(0)(1-z^2/h^2)$, and the brush height $h$  is 
  $\displaystyle h=\frac{8}{3}<z>$. 
For the simulations discussed in the manuscript we select $n=40$, so that the evaluated
$\rho(z)$ is  in between the step and the parabolic shapes (see Fig.~\ref{density_prf_graft1.5}).
In the following, we define $h$ following the definition of a step distribution, e.g.
\be \label{brush_height}
h=2     \frac{ \int_0^{\infty} z \rho(z)  dz}{\int_0^{\infty}  \rho(z)  dz}
\ee

\subsection{2.1 Brush compression properties}
\begin{figure}[ht]
\centering%
\includegraphics[width=8cm]{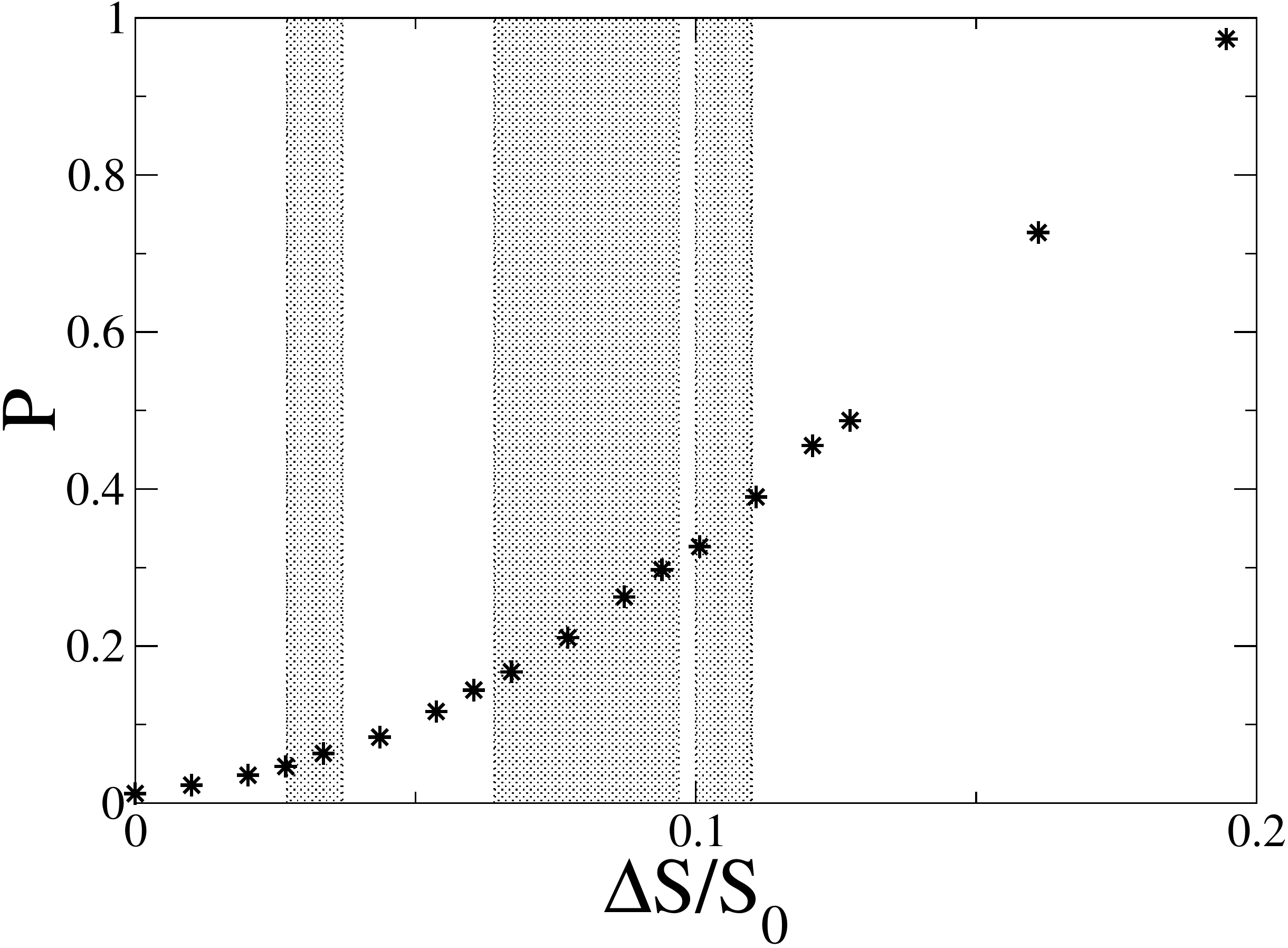}
\caption{%\baselineskip=12pt
The brush elastic response stored in the Young's modulus $E$. The modulus is defined as the coefficient between the relative brush compression and the applied pressure needed to produce it. The shaded areas indicate the ranges of brush surface compression corresponding to the three regions of flow inversion.}
\label{young}
\end{figure}

From simulations at equilibrium ($A=0$) in which the brush is compressed by a flat surface kept at a fixed position $S_{fix}$ we have estimated the Young's modulus $E$ as $\displaystyle E=\frac{P}{\Delta S/S_0}$. In this expression, 
$\Delta S/S_0=(S_{fix}-S_0)/S_0$ is the compression of the brush relative to the value $S_0$ indicating the brush surface without any imposed confinement and $P$ is the pressure exerted by polymers on the pressing apparatus.
In Fig.~\ref{young} we observe that $E$ is 
%almost linearly 
increasing with the compression.
We now notice that flow produces a compression of the brush (see Fig.~\ref{compression}).
Equating the compression imposed at equilibrium with the one produced by the applied flow (shaded areas in Fig.~\ref{young}) we conclude that the brush elastic response is modified by a change in $Wi$. 

\section{3. The Weissenberg number}
\begin{table}[t] 
\centering
Conversion $A\leftrightarrow Wi$
\vskip .2cm
\begin{tabular}{c | c || c | c || c | c || c | c} 
$A$ & $Wi$ & $A$ & $Wi$ & $A$ & $Wi$ & $A$ & $Wi$ \\
\hline
$0.005$ & $45$ & $0.022$ & $208$ & $0.047$ & $409$ & $0.064$ & $612$ \\
$0.007$ & $64$ & $0.023$ & $218$ & $0.048$ & $413$ & $0.065$ & $616$ \\
$0.01$ & $91$ & $0.028$ & $275$ & $0.05$ & $418$ & $0.066$ & $627$ \\
$0.015$ & $137$ & $0.03$ & $297$ & $0.051$ & $420$ & $0.067$ & $642$ \\
$0.016$ & $147$ & $0.034$ & $340$ & $0.055$ & $430$ & $0.07$ & $737$ \\
$0.017$ & $158$ & $0.037$ & $375$ & $0.0565$ & $588$ & $0.075$ & $788$ \\
$0.018$ & $167$ & $0.04$ & $400$ & $0.575$ & $593$ & $0.08$ & $832$ \\
$0.019$ & $175$ & $0.043$ & $405$ & $0.058$ & $595$ & $0.1$ & $1044$ \\
$0.02$ & $187$ & $0.045$ & $406.7$ & $0.06$ & $599$ \\
$0.021$ & $200$ & $0.046$ & $407.4$ & $0.063$ & $607$ \\
\end{tabular}
\caption{%\baselineskip=12pt
Tables of conversions between the constant acceleration $A$, imposed on each solvent particle to obtain a parabolic velocity profile, and the estimated a-dimensional Weissenberg number $Wi$.}
\label{conversioni}
\end{table}

\begin{figure}[ht]
\centering%
\includegraphics[width=9cm]{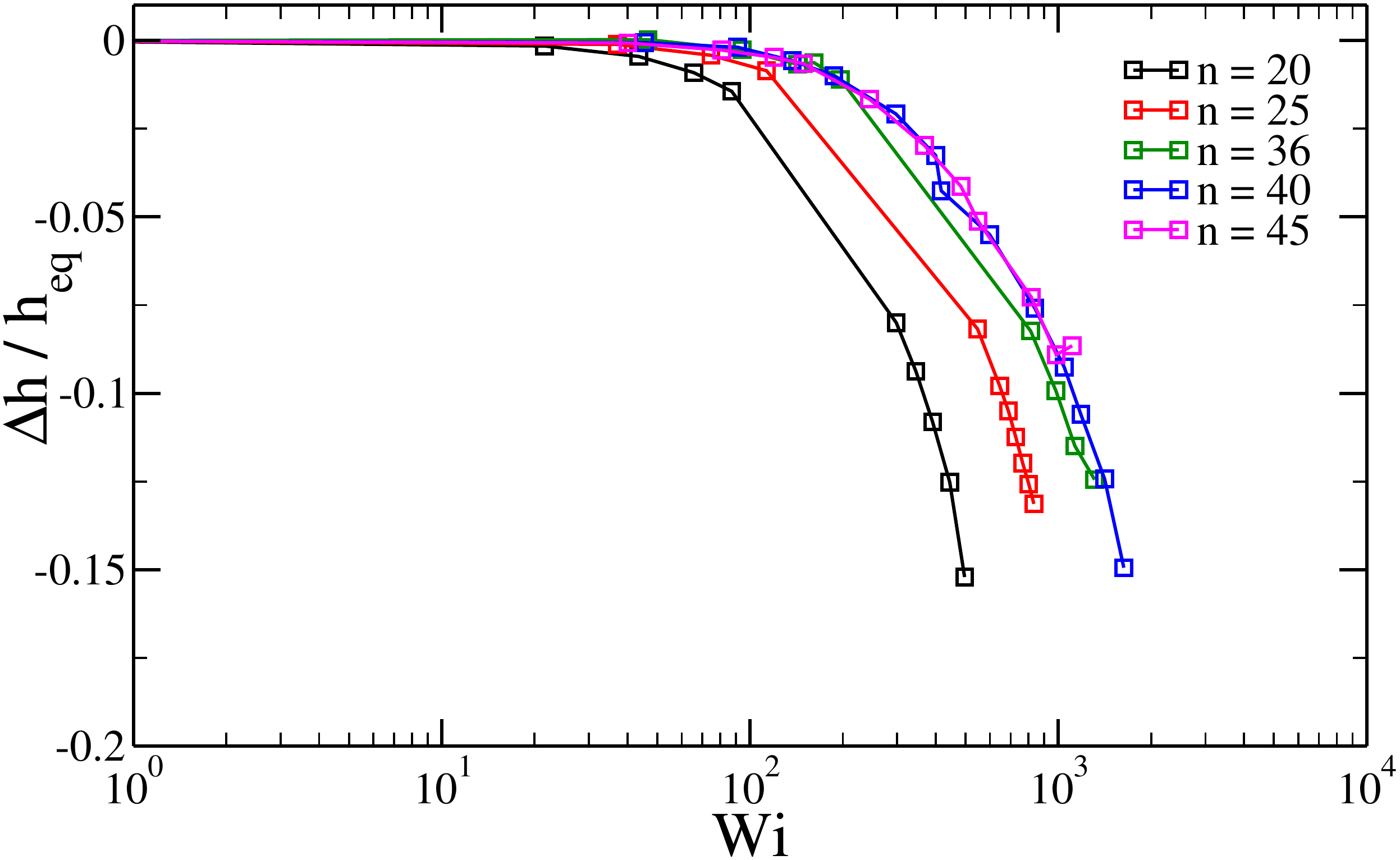}
\caption{%\baselineskip=12pt
The brush height $h$ relative to the equilibrium value $h_{eq}$ as function of $Wi$. The brush height is defined in section 2. For $Wi<100$ the height keeps basically constant, while for $Wi>100$ the flow has the effect to compress the brush.}
\label{compression}
\end{figure}
Instead of using $A$ as a measure of the flow intensity we prefer to introduce an adimensional quantity, the Weissenberg number $\displaystyle Wi:= t_{brush}/t_{flow}$, where
$t_{brush}$ should be a characteristic structural time of the unperturbed brush and $t_{flow}$ should provide 
an estimate of the typical time scale associated to the flow. 
We evaluate $t_{brush}$ as the relaxation time obtained by the decay of the autocorrelation function, averaged over all distinct polymers, of the end-to-end vector amplitude $R_{ee}= |\vec{r}_{0}-\vec{r}_n|$,
where $\vec{r}_{n}$ and $\vec{r}_{0}$ are the position vectors of, respectively, the last particle and the anchor of one chain inside the brush.
According to such definition, the relaxation time $t_{brush}$ assumes a dependency on the grafting density and on the degree of polymerization, thus representing a time scale of the whole brush rather than an isolated single-chain property.
%\footnote{inserire anche un grafico del decadimento e del fit con stretched exponential.}
The $R_{ee}$ auto-correlation function can be well represented by a stretched exponential decay 
$R_{ee} \sim e^{-(t/t_o)^\beta}$, where $t_o$ is the characteristic time scale and $\beta$ the so-called stretching 
exponent. Thus $t_{brush} = \Gamma[1/\beta] t_o/\beta$. For $\sigma_{graft}=1.5$ and $n=40$ we obtain $t_{brush}=1746$.
Regarding $t_{flow}$, we choose the inverse of the averaged shear rate $\dot{\gamma}$ calculated as
the ratio between the  maximum flow velocity $v_{max}$ in the channel and the corresponding z-coordinate $R$.
%  between the center of the parabolic flow and the location of the brush height $h$.
%\footnote{To be noted that the choice $\displaystyle \dot{\gamma}=t_{flow}^{-1}=\frac{v_{max}}{R_{eff}}$ underestimates the shear rate affecting the brush surface.}
Since our brush under flow experiences compression (see Fig.~\ref{compression}) and polymers are set only at the bottom wall, both $v_{max}$ and $R$ depend on the imposed acceleration $A$.

In table Tab.~\ref{conversioni} we list the correspondences between each value of $A$ and its $Wi$.

\section{4. Videos}
We show in movies SI1 and SI2 the dynamics of a single chain inside the brush for two different $Wi$, one giving flow inversion (movie~SI1) and the other with no flow inversion (movie~SI2): the recursive motion is constituted by an alternation of elongation-stretching-recoiling in both cases, thus it cannot be used as sufficient explanation for the backflow phenomenon.
%\movie[options]{placeholder box}{nome}

The motion of the whole brush under flow has been recorded in movies SI3 and SI4. In the first case no collective motion is detectable and no backward flow is observed in the velocity profile, while in the second case there is concurrence of a surface wave and of flow inversion.

\section{5. Finite size effect}
\begin{figure}[ht]
\centering%
\includegraphics[width=8cm]{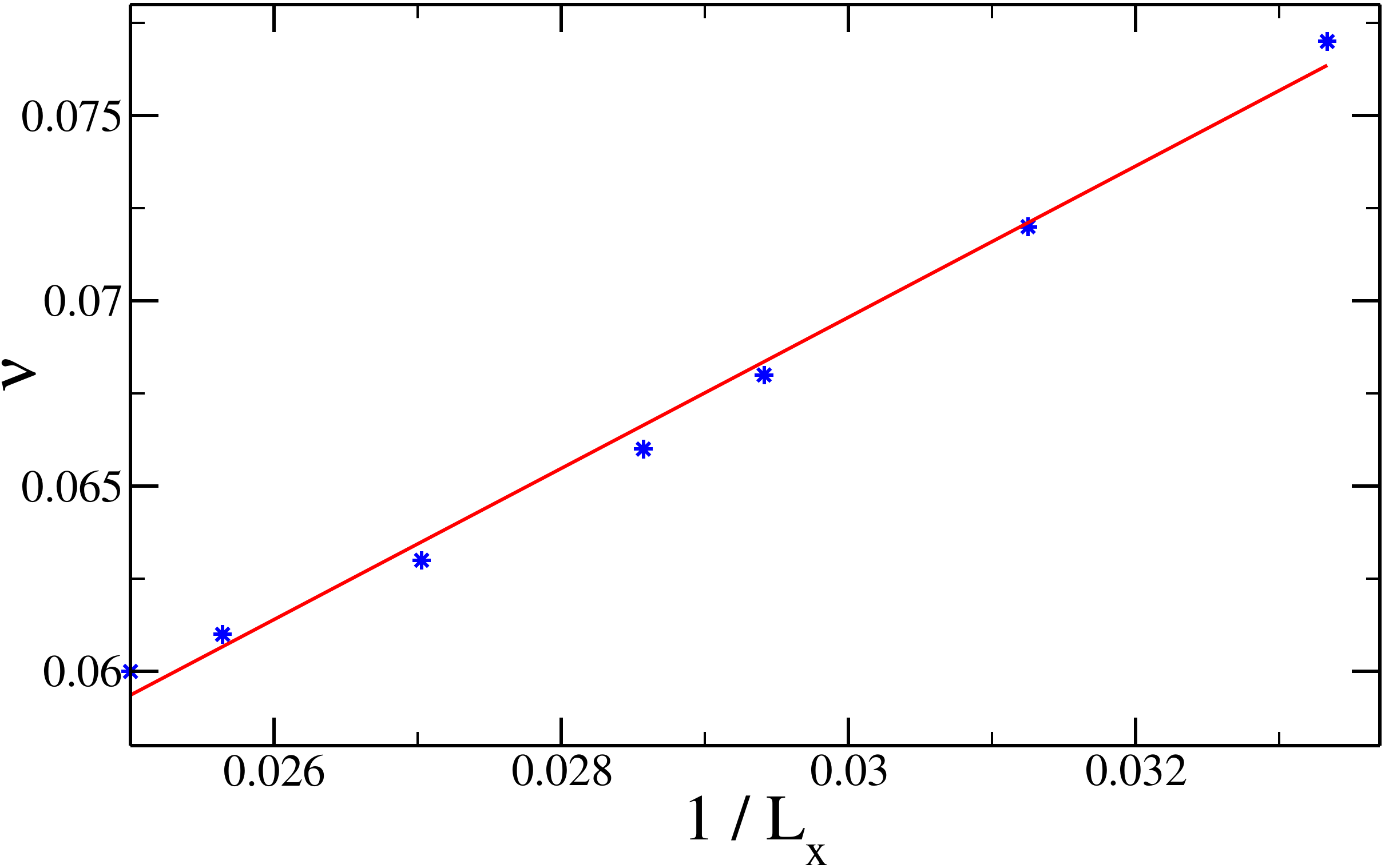}
\caption{%\baselineskip=12pt
Increasing the channel length $L_x$, $30<L_x<40$, at fixed acceleration $A=0.05$  the wave increases in wavelength ($\lambda =L_x$) and decreases in frequency such that the propagation speed keeps constant.}
\label{finite_size}
\end{figure}

We have investigated the influence of finite size effect increasing the channel length $L_x$ at fixed acceleration $A$: the wave persists. For a larger channel, $30 < L_x < 40$, the wave increases in wavelength ($\lambda =L_x$) and decreases in frequency, as shown in Fig.~\ref{finite_size}, such that the propagation speed keeps constant. The wave amplitude decreases and, in accordance with the Taylor's relation (see Letter, Eq.~(1)), $v_{min}$ decreases too.

Another test has been attempted increasing, up till the double, both $L_x$ and $L_z$ ($L_x=60$ and $L_z=100$). For such a system we found again, for $Wi=464$, a surface wave, with double amplitude and double wavelength (see movie~SI5).
However, investigating channels up to $L_x=360$, $L_y=20$ and $L_z=150$, we found that the amplitude of the surface wave 
does not keep on growing upon increasing $L_x$: in fact, its value seems to reach a plateau.  We interpret
this non-linear behaviour as a consequence of the finite extensibility of the polymers, e.g. due to 
the higher and higher energy required to significantly compress and stretch the brush.

\bibliography{biblio_wave}

\end{document}